\documentclass[prl,twocolumn,floatfix,showpacs,preprintnumbers]{revtex4-1}

\usepackage[latin2]{inputenc}
\usepackage{amsmath}
\usepackage{amssymb}
\usepackage{epsfig}
\usepackage{bm}
\usepackage{xcolor}
\usepackage{stackengine}
\usepackage{textcomp}
\usepackage{fontenc}

\DeclareMathOperator{\sgn}{sgn}

\begin{document}

\title{Self-organized lane-formation in bidirectional transport of molecular motors}

\author{Robin Jose}
\author{Ludger Santen}
\affiliation{Department of Theoretical Physics $\&$ Center for Biophysics, Saarland University, 66123 Saarbr\"ucken, Germany}

\date{\today}

\begin{abstract}
\noindent {\bf ABSTRACT}  
Within cells, vesicles and proteins are actively transported several micrometers along the cytoskeletal filaments. The transport along microtubules is propelled by dynein and kinesin 
motors, which carry the cargo in opposite directions. Bidirectional intracellular transport is performed with great efficiency, even under strong confinement, as for example in the axon.
For this kind of transport system, one would expect generically cluster formation. In this work, we discuss the effect of the recently observed self-enhanced binding-affinity along the kinesin trajectories on the MT. We introduce a stochastic lattice-gas 
model, where the enhanced binding affinity is realized via a floor-field. From Monte Carlo simulations and a mean-field analysis we show that this mechanism can lead to self-organized symmetry-breaking and lane-formation which indeed leads to efficient bidirectional transport in narrow environments.
\end{abstract}
 
\maketitle

\noindent {\bf INTRODUCTION}\\
\noindent The efficiency of intracellular transport is one of the most intriguing features of biological cells. Different kinds of cellular cargo have to be transported to specific locations in 
order to maintain the cells' functionality. Intracellular transport can be driven by molecular motors, i.e. specialized  proteins that can carry cargo along polar filaments of the cytoskeleton  \cite{Hirokawa2005,Brown2003,Herold2012,Hollenbeck2005}. Molecular motors, such as the microtubule (MT) 
associated proteins (MAPS) kinesin and dynein, step stochastically along MTs in a given preferred direction: Kinesins step typically toward the plus-end and dyneins to the minus-end. Molecular 
motors are able to carry big (on the scale of the cell) objects through crowded environments.

We focus on bidirectional motor-driven transport under spatial confinement, which is for example relevant for intracellular transport in axons. In this kind 
of environment, active transport is particularly difficult to organize, since cluster formation is generically observed in spatially extended one-dimensional 
systems  \cite{Ebbinghaus2011,Appert-Rolland2015, Evans1995, Korniss1999, Jose2020}. Clusters can either have stationary particle output \cite{Evans1995} or can lead to long times of blockages such as for non-Markovian site-exchange \cite{Jose2020}. The general question we address in this work is the following: How do confined systems of active particles self-organize to realize efficient bidirectional transport states? 

\begin{figure}[b!]
	\centering
	\includegraphics[width=0.48\textwidth]{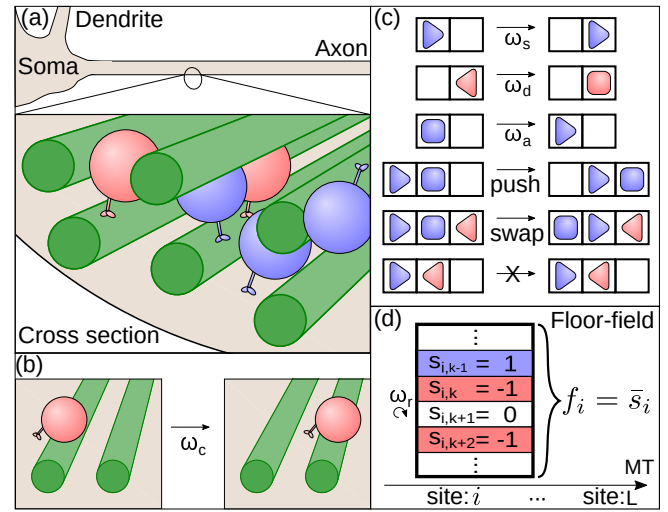}
	\caption{\textbf{(a)} Scheme of a neuron, indicating the crowded environment and confinement inside axons, including MTs, plus-particles (blue) and minus-particles (red). \textbf{(b)} Unbound particles switch filaments with rate $\omega_c$. \textbf{(c)} Particle dynamics in the exclusion process. Triangles mimic bound particles, the tip indicates the direction. Unbound particles are shown by squares. Bound particles can step or detach, unbound particles can reattach. If a particle attempts to step onto a site occupied by an unbound particle, it can either push it away or swap position. Two bound particles block each other via exclusion. \textbf{(d)} The floor-field state $f_i$ is averaged over all sub-states for every site $i$ of the lattice.}
	\label{Fig1}
\end{figure}

Motor-driven transport has been described by variants of the totally asymmetric exclusion processes (TASEP) which combine the 
directed stochastic motion of particles on a one-dimensional lattice with hard-core exclusion and Langmuir-kinetics \cite{Parmeggiani2004, Ebbinghaus2011, Pinkoviezky2017}. 
In principle, the particle exchange with a reservoir would allow for bidirectional transport, in case of large diffusivity of unbound particles. However, if the unbound particles 
are localized, so far no mechanism has been suggested which leads to efficient bidirectional transport. A rather direct approach is the self-organization in sub-systems each of which carries unidirectional transport.  A recent hypothesis is that posttranslational modifications on MTs might organize transport in neurons \cite{Gadadhar2017, Janke2014}. This kind of organization has been observed for example in dendrites, where the MTs are oppositely oriented \cite{Tas2017} and in MT doublets in cilia \cite{Stepanek2016}. Furthermore, motor proteins can regulate MTs themselves \cite{Johann2012}, and MT-dynamics \cite{Ebbinghaus2011} and tau \cite{Ebneth1998} can affect motor transport.

Recent experimental findings suggest a possible mechanism leading to efficient bidirectional transport on MT bundles where no  {\it a priori} compartmentalization exits. 
Shima \textit{et al.} \cite{Shima2018} reported 
that binding affinity of MTs for kinesin motors is self-enhanced along the kinesin trajectories which modify the MTs. This kind of self-induced preferential binding can be understood 
as a true realization of a floor-field, which has been successfully introduced as a virtual mechanism in order to generate e.g. lane-formation in bidirectional pedestrian flows 
\cite{Burstedde2001, Nowak2012, Chraibi2019, Gibelli2019,Schadschneider2010}.

In this paper, the transport problem is formulated as a TASEP with Langmuir-kinetics, where we additionally consider an explicit particle reservoir and a floor-field. Our theoretical model describes the key features of bidirectional axonal transport but considerably reduces the complexity of the biological reference system. \\

\noindent {\bf THE MODEL}\\
We study a TASEP with Langmuir-kinetics of two particle species moving on a pair of parallel, identically polarized one-dimensional filaments. 
The model filaments (MTs) are represented as one-dimensional, static lattices. Lattice sites can either be empty or occupied by a single particle. 
We consider two types of particles, i.e. moving to the plus-end of the filament ($\tau=1$, blue in Fig.\,\ref{Fig1}) and to the minus-end ($\tau=-1$, red). 

\textit{Particle dynamics:} Both types of particles can either be bound or unbound to a filament (triangles or squares in Fig.\,\ref{Fig1}(c)). Bound particles step to the neighboring
site (target-site) with rate $\omega_s$  or detach from the filament with rate $\omega_d$ (Fig.\,\ref{Fig1}(c)). In order to study lane-formation as 
a bulk effect, we are considering periodic boundary conditions.
Particles which detach from the filament stay at the same lattice-site, unlike in typical models with Langmuir-kinetics where particles move to a bulk reservoir \cite{Parmeggiani2004, Ebbinghaus2011}. This 
feature is crucial for modeling transport in crowded environments, where unbound particles cannot simply diffuse away from clusters. 

Unbound particles can reattach to the filament with rate $\omega_a$ or change to an unbound state on the other filament with a coupling rate $\omega_c$ (Fig.\,\ref{Fig1}(b)),
where the position is kept.
Particles interact with each other via hardcore repulsion (Fig.\,\ref{Fig1}(c) bottom). 
For a particle which is selected to step we distinguish three cases. $(i)$ If the target-site is free, the step is executed. $(ii)$ If the target-site is occupied by a bound particle the step is rejected. $(iii)$ If the target-site is occupied by an unbound particle, the unbound particle is either pushed to next site (in moving direction of the stepping particle) or exchanges position with it (swapping). If both pushing and swapping are possible, one of the two possibilities is selected with probability $1/2$. If the site in moving direction next to the unbound particle is occupied, swapping is executed. (Fig.\,\ref{Fig1}(c)). 

\textit{Floor-field dynamics:} In \cite{Shima2018, Peet2018} an axial elongation of the MT by kinesin has been reported. The elongation is related to a meta-stable tubulin-state which has a higher binding affinity for kinesins. This effect is implemented via a floor-field which considers the number of MT protofilaments, $N_p=13$. A floor-field $f_i$ is assigned to each lattice-site $i$, which is given by
\begin{equation}
f_i = \frac{1}{N_p} \sum_{k=1}^{N_{p}} s_{i,k}
\label{Eq:Floorfield}
\end{equation}
where $k$ denotes the index of the protofilament which is permanently assigned to the particles until they detach from the (proto-)filament. Therefore, $f_i$ represents the average of $N_p$ sub-states $s_{i,k} \in \{ -1, 0, 1 \}$. The value of $f_i$
is updated if particle steps to site $i$ and thereby sets the value of a given sub-state $s_{i,k}$ to $+1(-1)$ in case of $+(-)$ directed motors.
The sub-state can decay back to 0 again with rate $\omega_r$ (Fig.\,\ref{Fig1}(d)). Averaging over $N_p$ sub-states introduces a memory effect which stabilizes the preferential adsorption of a given type of particle, i.e. the amplitude of the floor-field determines the robustness of the floor-field against changes of the affinity by single oppositely directed particles. The sub-division of the floor-field into "protofilaments" is also consistent with the observation that low kinesin concentration may lead to a curvature of MTs which signifies a coexistence of excited and non-exited tubulin states (\cite{Peet2018}).

The state $f_i$ influences the binding affinity of particles $\omega_{a,i}$ given by
\begin{equation}
\omega_{a,i}=
\begin{cases}
\displaystyle \omega_a^0\,\mu^{\lvert f_i \rvert} ,  
& \;\;\;\; \tau = \sgn\left(f\right), \vspace{1mm}\\
\displaystyle \omega_a^0\,\frac{1}{\mu^{\lvert f_i \rvert}},  
& \;\;\;\; \tau \neq \sgn\left(f\right), \vspace{1mm}
\end{cases}
\label{Eq:modification}
\end{equation}
where  $\omega_a^0$ is the free attachment rate and $\mu \geq 1$ is called affinity modification factor. This modification leads to higher binding rates if the floor-field state $f_i$ was predominantly set by particles of the same 
type $\tau$ as well as lower rates for opposing combinations. If $\mu = 1$ or $f_i=0$, the interaction is neutral. We consider a symmetric excitation for dynein and kinesin motors, though so far experimental evidence for a modification of the MT-structure by dynein is still lacking.  \\

\noindent {\bf RESULTS AND DISCUSSION}\\
We study the influence of the floor-field on the particle flux $J$ as a measure of transport efficiency as well as symmetry-breaking and self-organized lane-formation. First, we introduce a mean-field analysis and then compare results to Monte Carlo (MC) simulations.

\textit{Mean-field analysis:} As a reference, we consider TASEP models \cite{Derrida1998}, two-species, bidirectional exclusion processes \cite{Evans1995, Arndt1998, Jose2020}, as well as combinations of TASEP and Langmuir-kinetics \cite{Parmeggiani2004, Ebbinghaus2009, Ebbinghaus2011}. From these models, a mean-field estimation \cite{Evans2003} of the flux $J_{\text{ud}}=\rho_{\text{eff}}(1-\rho_{\text{eff}})$ can be deduced for unidirectional one-filament systems with Langmuir-kinetics (\cite{suppl}). We use $J_{\text{ud}}$ for judging on the transport efficiency. Note that fluxes are scaled by $\omega_s^{-1}$ and the system size $L$.

To include the floor-field dependency in a mean-field model, we assume a simplified unbound state (\textit{u}) shared for both filaments called top (\textit{t}) and bottom (\textit{b}). The average floor-field $f$ is represented by the normalized difference in particle densities $\Delta_{t,+} =(\rho_t^+-\rho_t^-)/\rho^+$ for the plus-species and the top filament (bottom analog) so that we can formulate the mean-field equations exemplary for plus-particles (details in the supplemental material \cite{suppl})
\begin{align}
	\frac{\partial \rho_t^+}{\partial t}  = \ & \omega_a^0 \mu^{\Delta_{t,+}}\rho_u^+ - \omega_d \rho_t^+  \nonumber\\	
	\frac{\partial \rho_u^+}{\partial t} = \ & \omega_d\left(\rho_t^++ \rho_b^+ \right) -\left( \mu^{\Delta_{t,+}} +  \mu^{\Delta_{b,+}}\right)\omega_a^0 \rho_u^+ \\
	\frac{\partial \rho_b^+}{\partial t} = \ & \omega_a^0\mu^{\Delta_{b,+}}\rho_u^+ - \omega_d \rho_b^+ . \nonumber	
\end{align}
Additionally, we get the identity $\rho^{\pm} = \rho_t^{\pm} + \rho_b^{\pm} + \rho_u^{\pm}$ from particle conservation. In the stationary state, we find the equation for the difference in densities on the top filament $\Delta$ as
\begin{align}
\Delta  = \frac{\left(\mu^{\Delta} - \mu^{-\Delta} \right)}{\frac{1}{\omega_d}\left(\mu^{\Delta} + \mu^{-\Delta} \right) + \frac{\omega_d}{\omega_a^0}}  . \label{Eq:mean-field-Delta2}
\end{align}
Eq.\,\ref{Eq:mean-field-Delta2} is numerically solvable and shows a pitchfork bifurcation, at a critical $\mu= \mu_{crit}$: For $\mu <\mu_{crit}$ eq.\,\ref{Eq:mean-field-Delta2} has only a single solution given by $\Delta_0=0$, while for  $\mu > \mu_{crit}$ the solution $\Delta_0=0$ gets unstable and two stable points at $\Delta_{\pm}$, depending on $\omega_d$ and $\omega_a^0$, occur. We also find that the floor-field has to modify the affinity for both species, otherwise only a symmetric solution can be found \cite{suppl}.

By solving Eq.\,\ref{Eq:mean-field-Delta2}, the flux is estimated by 
\begin{align}
J_{\text{MF}} = \ & \rho_t^+\left(1-(\rho_t^++\rho_t^-)\right). \label{Eq:mean-field-flux}
\end{align}

\begin{figure}[t!]
	\centering
	\includegraphics[width=0.48\textwidth]{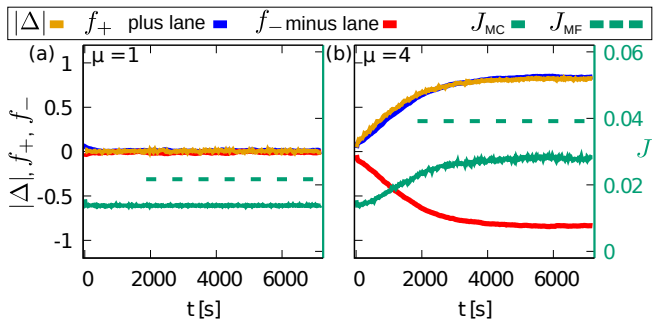}
	\caption{Time dependence of the flux and particle densities for a system of $L=1000$, $\rho=0.05$, and random initial configurations. Values of the floor-fields $ f_+$, $f_-$ and difference in densities $\Delta$ are given at the left axis; values of the total flux $J$, and the mean-field estimation $J_\text{MF}$ are given at the right axis. (a)  $\mu = 1$, (b) $\mu = 4$.  }
	\label{Fig2}
\end{figure}
\textit{Parameters:} We used the experimental results of \cite{Shima2018} to select the relevant parameters of the model, given 
in table\,1 in the supplemental material \cite{suppl}. 
We kept the rates $\omega_s$, $\omega_d$, $\omega_a^0$ and $\omega_r$ constant. The relevant density regime is rather difficult to estimate. On the one hand 
the fraction of occupied binding site is rather low. On the other hand molecular motors carry rather big objects (20 nm and 50 nm for axonal vesicles 
\cite{Harris1995,Hall1991}, compared to 8 nm step-size for most kinesin and dynein motors \cite{Svoboda1993, Hirakawa2000}) such that the density in terms of the 
occupied volume along the MT 
is considerably higher. Therefore, we did not focus on the low density regime of $\rho \approx 0.01$, which has been addressed in \cite{Shima2018} but varied the particle density in order to study the stability of the bidirectional transport in our model. The chosen lengths of approximately 1000 sites, which correspond to MTs of length 8 $ \mu$m, is in accordance to the typical MT-length in axons \cite{Shaebani2016, Yu1994}. 
The range of the affinity modification $\mu$ is motivated by different experiments in which kinesin binding affinity has been measured for different types of 
MTs. In \cite{Shima2018, Morikawa2015}, GTP-MTs show three to four times higher affinity than GDP-MTs and comparing \cite{Rosenfeld1996} with \cite{Nakata2011}, the 
affinity is five times higher. The choice of coupling rates, filament number and the number of sub-states in the floor-field implementation is discussed in the supplemental material \cite{suppl}.

\textit{MC-Simulations:} We investigate the influence of the floor-field on our stochastic model by performing MC-simulations with two filaments started with neutral floor-fields and randomly distributed 
particles. The total particle density is given by $\rho_{tot} = \rho_+ + \rho_- = 2 \rho_+$.

A time-evolution of the system is shown in Fig.\,\ref{Fig2} averaged over 100 simulations. Yellow lines show the difference in densities $\Delta$. A filament with average floor-field $ f=1/L\sum_{i=1}^L f_i > 0$ is called plus-lane and $ f < 0$ minus-lane. The floor-field $ f_+$ ($ f_-$) of the plus (minus)-lane is shown in blue (red), and the total flux $J$ in green (right axis).

Without modification, i.e. $\mu=1$ in panel (a), no symmetry-breaking is observed. There is no significant difference between $f_+$ and $f_-$, and particles are distributed equally ($\Delta=0$). By raising $\mu$, the floor-field values split up and $\Delta$ increases. For $\mu=4$, $ f_+$ ($ f_-$) and $\Delta$ almost reach the extreme values $\pm 1$, meaning a quasi separation of particles and totally asymmetric floor-fields. This lane-formation is stable and the time-evolution shows very little sample to sample fluctuations. Also the difference in the particle distribution $\Delta$ is in good agreement with the average floor-field $|f|$ which makes $\Delta$ a good representation for $f$ in the mean-field analysis.

The stationary flux (green) increases for higher $\mu$ when the floor-field is stabilized (Fig.\,\ref{Fig2}(b) with $\mu=4$). In case of $\mu=4$ ($\mu=6$) an average effective velocity of $\approx$ 270 nm/s (350 nm/s) for a motor protein whereas the free stepping velocity of bound kinesins is presumed to be 480 nm/s in this work (\cite{Shima2018}). As expected the mean-field solution (dashed green line in Fig.\,\ref{Fig2}) overestimates the flux considerably, since a homogeneous distribution of particles is assumed, while in the full model there are strong density-correlations due to cluster formation. However, the initially symmetric two-lane system spontaneously breaks symmetry so both lanes carry stationary and oppositely directed net flows. 

\begin{figure}[t]
	\centering
	\includegraphics[width=0.48\textwidth]{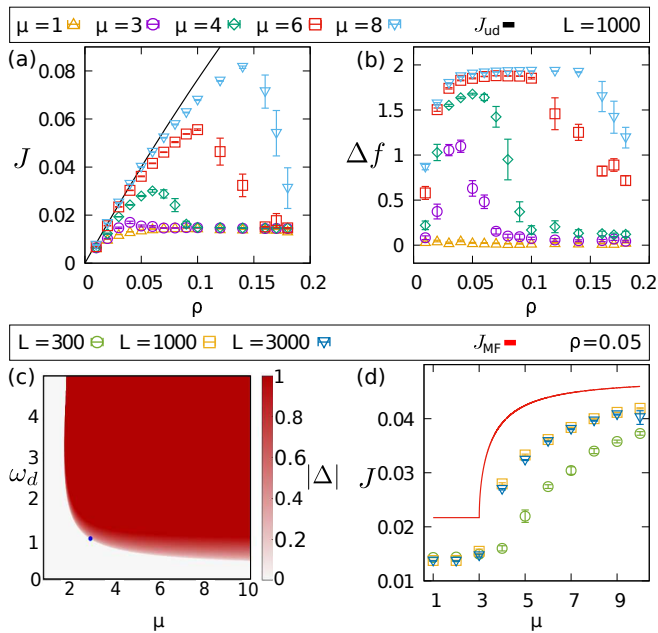}
	\caption{Transport efficiency \textbf{(a)} and symmetry-breaking ( \textbf{(b)}, \textbf{(d)}) under variation of density and affinity modification. Simulations did run for 3 hours real-time, measurements started after 1 hour. Panel \textbf{(c)} shows the phase-space for symmetric and asymmetric solutions in the mean-field model.}
	\label{Fig3}
\end{figure}

In Fig.\,\ref{Fig3}(a), we show the density dependence of the flux for different $\mu$ in comparison to the unidirectional flux. Simulation results show that the transport efficiency of the system is significantly increased for ($\mu \geq 3$) compared to the plateau obtained without floor-field ($\mu = 1$). Actually, the flux reaches almost the value of the corresponding unidirectional flux to $J_\text{ud}$ until it breaks down to the traffic jam plateau value, similar for all $\mu$. The density at which the transition to the plateau value is observed, depends on $\mu$. Note that the stationary state is not always reached at high densities if we initialize the system with a random configuration, indicated by the larger error bars in the high density regime caused by meta-stable clusters (Fig.\,\ref{Fig3}(a)), which have not been dissolved within the simulation time.

Lane-formation is well characterized by the difference in floor-field $\Delta f= f_+  -  f_- $ measuring asymmetry between filaments and is shown in Fig.\,\ref{Fig3}(b). The base line corresponds to symmetric fields without lane-formation for $\mu = 1$. By increasing $\mu$, the asymmetry develops in a density dependent range before $\Delta f$ drops down. Results of panel (a) and (b) indicate a lane-formation and quasi ordering the system into two sub-systems with oppositely directed flux. When the self-organization breaks down, traffic jams are forming on both lanes and transport efficiency is not enhanced anymore. This is consistent with lane-formation observed in other floor-field models \cite{Burstedde2001}.

The influence of $\mu$ on the symmetry-breaking is further examined in Fig.\,\ref{Fig3}(c) and (d) by comparing mean-field results to MC-simulations. In panel (c), a phase diagram from mean-field analysis for $|\Delta|$ under variation of $\mu$ and $\omega_d$ is shown for fixed $\omega_a^0=5s^{-1}$. The blue dot marks $\mu_{crit}$ for $\omega_d$ used in simulations and agrees with Fig.\,\ref{Fig2} and Fig.\,\ref{Fig3} (b). The border of $|\Delta|>0$ shows that $\mu_{crit}>1$ for arbitrary $\omega_d$. There is only a small region where $0<|\Delta|<1$ because the mathematical solution of Eq.\,\ref{Eq:mean-field-Delta2} can be larger than the physical border of $|\Delta|=1$, hence particles are completely separated. The transition is sharper for shorter run lengths (larger $\omega_d$). In panel (d), $J$ is growing under variation of $\mu$ for constant $\rho=0.05$ and different $L$. Remarkably, the transition from a symmetric to a stable asymmetric solution is captured by the mean-field approach and even the predicted value $\mu_{crit}$ agrees well with simulation results. The transition is sharper for $L\geq 1000$ than for $L=300$, hence, the larger system is better approximated by the mean-field model. Also, larger systems have higher fluxes. This is in contrast to the plateau value for $\mu =1$ which decreases with the system size. For even larger $L$ it is computationally hard to achieve stationary states but we expect the system to still self-organize in lanes due to stable lanes if already started in such conditions (supplemental material \cite{suppl}).\\

\noindent {\bf CONCLUSION AND OUTLOOK}\\
To summarize, we introduced a stable mechanism for efficient bidirectional transport of active particles in one-dimensional systems under strong confinement. 
This mechanism is based on self-organized lane-formation. Directed lanes may be predefined in engineered systems, however, this is not always the case for transport of animals or humans as for instance in pedestrian dynamics where self-organized lane-formation occurs \cite{Burstedde2001, Nowak2012, Chraibi2019, Gibelli2019,Schadschneider2010}. The influence of the floor-field on particle binding
was inspired by recent experimental results on self-induced strengthening of the kinesin MT-affinity, but could 
also be realized by other modifications of MTs. Lane-formation can be captured by a mean-field approach, which shows the mechanism is stable against local density fluctuations.

The stability of lane-formation is remarkable in several respects. First of all, lane-formation is observed in the biologically relevant low density regime. This is in contrast to other mechanisms, based on particle-particle interactions \cite{Klumpp2004}, which lead to symmetry-breaking at high densities 
and therefore low particle velocities, while \textit{in vivo} observations of e.g. axonal vesicle transport show that vesicles transported by molecular motors reach the free 
stepping velocities of kinesin. Second, we observe the coexistence of transport in both directions on a coupled pair of filaments, which goes 
beyond symmetry-breaking mechanisms reported as discussed in e.g. \cite{Popkov2001} where symmetry-breaking leads to unidirectional  transport.  
Third, our model describes the low mobility of unbound particles, which may trigger cluster formation in bidirectional transport and illustrates the 
stability of the suggested mechanism. From our point of view, our results indicate that stable bidirectional flows are more easily 
realized by modifications of the filaments rather than interactions between particles. 

The importance of the MT structure on transport has recently been pointed out \cite{Janke2014, Gadadhar2017, Tas2017, Stepanek2016}. Bidirectional intracellular transport 
is organized on oppositely oriented filament bundles in dendrites \cite{Tas2017} and on parallel oriented MT doublets in cilia \cite{Stepanek2016}. In axons, 
however, so far a similar organization of the MT network has not been identified. Our findings indicate that the posttranslational modification by motors and self-induced preferential binding of 
one or the other motor species could indeed lead to stable bidirectional 
transport in an {\it a priory} unipolar MT network. A self-induced amplification of the binding affinity must be given for both particle species. 
Otherwise, the density of oppositely oriented particles on the same filament is too high to realize efficient transport states. 

Concerning the robustness and efficiency of the proposed 
lane-formation in our model for intracellular transport, 
it would be of great interest to obtain further insight to the 
interplay between dynein and kinesin motors, microtubules and MAPS, which might have a strong impact on the (self-)organization of 
intracellular transport. 
\\

\begin{acknowledgments}
\noindent {\bf ACKNOWLEDGEMENTS}\\
\noindent This work was funded by the Deutsche Forschungsgemeinschaft (DFG) through Collaborative Research Center SFB 1027 (Project A8).\\
\end{acknowledgments}

%
%
%



\end{document}


\title{Self-organized lane-formation in bidirectional transport of molecular motors\\
	SUPPLEMENTAL MATERIAL}

\author{Robin Jose}
\author{Ludger Santen}
\affiliation{Department of Theoretical Physics $\&$ Center for Biophysics, Saarland University, 66123 Saarbr\"ucken, Germany}

\date{\today}

\maketitle

\textit{Mean-field analysis:} 
	In the continuum limit for a single, unidirectional exclusion process with particle exchange to a reservoir, the mean-field approach is given by [29] 
	\begin{equation}
	\frac{\partial \rho (x,t)}{\partial t} = \frac{\partial J(x,t)}{\partial x} + \omega_a(1-\rho(x,t)) -\omega_d\rho(x,t).
	\label{Eq:Langmuir-tasep}
	\end{equation}
	Here, we consider a system with periodic boundary conditions and translational invariant initial conditions, such that 
	$\frac{\partial J}{\partial x}=0  $ and $\frac{\partial \rho}{\partial t}=0  $ holds in the stationary state. In absence of a floor-field, the effective density of particles bound to the filament is given by Langmuir kinetics [11, 29]. Hence, we obtain the following estimates for the stationary density and flux:
	\begin{equation}
	J_{\text{ud}}=\rho_{\text{eff}}(1-\rho_{\text{eff}}) \quad \quad \quad   \rho_{\text{eff}} = \rho \frac{\omega_{a}^0}{\omega_{a}^0+\omega_{d}}.
	\label{Eq:TASEP}
	\end{equation}	
	Next, we consider in a system consisting of two filaments (top and bottom) and a mutual reservoir (unbound) of infinite capacity. Please note that we consider a single particle reservoir in the mean-field approach instead of two weakly coupled reservoirs in the full model, each of which being coupled to one of the two filaments. This simplification is valid  because the weak coupling of the two particle-reservoirs of the full model suppresses the coupling of the density-fluctuations between the two filaments. Density-fluctuations, however, are not described by the mean-field approach. 
	
	
	Whereas the detachment rates  are  constant, the attachment rate is a function of particle densities 
	and  describes the impact of the floor-field in the full model.
	Eq.\,2 in the main text describes that the attachment
	depends on the floor-field $f_i$ on the particular site $i$ on which the particle wants to attach. Due to the translational invariance of the model we consider consistently an  average floor-field $f$. We estimate the average floor-field by the normalized difference of densities $\rho^+$ and $\rho^-$ on the given filament, i.e.  $f^t=(\rho_t^+-\rho_t^-)/\rho^+$ for a plus-particle on the top filament. The attachment rates for plus- and minus-particles on the top filament $ \omega_a^{t,\pm} = \omega_a^{t,\pm} (\rho_t^+,\rho_t^-)$  are then given by 
	\begin{align}
	\omega_a^{t,+}  = \ & \omega_a^0 \mu^{(\rho_t^+-\rho_t^-)/\rho^+}  \\	
	\omega_a^{t,-}  = \ &  \omega_a^0 \mu^{(\rho_t^--\rho_t^+)/\rho^-}.
	\label{Eq:mf_attachment}
	\end{align}
	The equations given above hold for attachment to the top-filament. An analogous set of equations describes the attachment to the bottom-filament.
	 
	As a result we arrive at the following mean-field equations for the two-filament system with particle reservoir and the previously defined attachment rates $\omega_a^{t,\pm}, \omega_a^{b,\pm}  $, which depend on the difference of plus and minus motor density: 
	\begin{align}
	\frac{\partial \rho_t^+}{\partial t}  = \ & \omega_a^{t,+} \rho_u^+ - \omega_d \rho_t^+ \\	
	\frac{\partial \rho_u^+}{\partial t} = \ & \omega_d\left(\rho_t^++ \rho_b^+ \right) \\   & -\left(\omega_a^{t,+} + \omega_a^{b,+} \right) \rho_u^+ \nonumber \\
	\frac{\partial \rho_b^+}{\partial t} = \ & \omega_a^{b,+} \rho_u^+ - \omega_d \rho_b^+ 	,
	\label{Eq:mean-field}
	\end{align}
where $ \rho_u^+$ denotes the density of the plus-particles in the particle reservoir. We get an analogous set of particles for minus particles. Furthermore, particle conservation leads to  the relation $\rho^+ = \rho_t^+ + \rho_b^+ + \rho_u^+$ for
plus-particles and  $\rho^- = \rho_t^- + \rho_b^- + \rho_u^-$ for minus-particles. For simplicity, we consider $\rho^+=\rho^-$.

We define $\Delta =(\rho_t^+-\rho_t^-)/\rho^+$ which we use as an estimate of the average floor-field $f$ (compare to Fig.\,2). We also make use of the symmetry between plus- and
	minus-particles, i.e. we choose $\rho_t^+=\rho_b^-$, $\rho_b^+=\rho_t^-$ and $\rho_u^+=\rho_u^-$ and drop the $\pm$ index. Using these assumptions and definitions 
	we get: 
	\begin{align}
	\frac{\partial \rho_t}{\partial t}  = \ & \omega_a^0 \mu^{\Delta} \rho_u - \omega_d \rho_t \label{Eq:mean-field-simple1}\\	
	\frac{\partial \rho_u}{\partial t} = \ & \omega_d\left(\rho_t + \rho_b \right)  -\left(\omega_a^0 \mu^{\Delta} + \omega_a^0 \mu^{-\Delta}\right) \rho_u \label{Eq:mean-field-simple2} \\
	\frac{\partial \rho_b}{\partial t} = \ & \omega_a^0 \mu^{-\Delta} \rho_u - \omega_d \rho_b.	
	\label{Eq:mean-field-simple3}
	\end{align}
	From Eq.\,\ref{Eq:mean-field-simple1} to Eq.\,\ref{Eq:mean-field-simple3}, we find the following  equation for $\Delta$:  
	\begin{align}
	\frac{1}{\omega_d} \frac{\partial \Delta}{\partial t}  = \frac{\omega_a^0 \ \rho_u}{\omega_d \ \rho} \left(\mu^{\Delta} - \mu^{-\Delta} \right) -\Delta . \label{Eq:mean-field-Delta1}
	\end{align}
	Making use of normalization and Eq.\,\ref{Eq:mean-field-simple2}, we can determine an equation for $\Delta$ in the stationary state
	\begin{align}
	\Delta  = \frac{\left(\mu^{\Delta} - \mu^{-\Delta} \right)}{\frac{1}{\omega_d}\left(\mu^{\Delta} + \mu^{-\Delta} \right) + \frac{\omega_d}{\omega_a^0}}  . \label{Eq:mean-field-Delta2}
	\end{align}
	Eq.\,\ref{Eq:mean-field-Delta2}  is numerically solvable and shows a pitchfork bifurcation at $\mu= \mu_{crit}(\omega_a^0, \omega_d)$ from a stable equilibrium point at $\Delta_0=0$ to the unstable equilibrium point $\Delta_0=0$ and two stable points at $\Delta_{\pm}$. Note that this asymmetrical solution can be found for $|\Delta|>1$ in some cases. By definition, the difference of physical densities cannot be larger than 1, so a solution $|\Delta|>1$ will correspond to a total separation of particles at the border of the definition of $\Delta$.

As a next step, we explicitly calculate the densities in each state by using the asymmetric solutions for $\Delta$ 
	\begin{align}
	\rho_t = \ & \frac{ \rho \ \Delta_{\pm} \ \mu^{\Delta_{\pm}}}{\left(\mu^{\Delta_{\pm}} - \mu^{-\Delta_{\pm}}\right) } \label{Eq:mean-field-solution1}\\	
	\rho_u = \ & \frac{\omega_d}{\omega_a^0}\frac{ \rho \ \Delta_{\pm} }{\left(\mu^{\Delta_{\pm}} - \mu^{-\Delta_{\pm}}\right) } \label{Eq:mean-field-solution2} \\
	\rho_b = \ & \frac{ \rho \ \Delta_{\pm} \ \mu^{-\Delta_{\pm}}}{\left(\mu^{\Delta_{\pm}} - \mu^{-\Delta_{\pm}}\right) }.	
	\label{Eq:mean-field-solution3}
	\end{align}
	Using these results, we can calculate the flux for plus-particles on the top lane 
	\begin{align}
	J_{\text{MF}}=J_t^+ = \ & \rho_t^+\left(1-(\rho_t^++\rho_t^-)\right) \label{Eq:mean-field-flux}
	\end{align}
	and analogously for the bottom lane as well as for minus-particles.

 \begin{table}[b]
 	\caption{Reference set of parameters. The rates $\omega_s$, $\omega_a$, $\omega_d$ and $\omega_r$ are extracted from [12]. }
 	\centering
 	\begin{tabular}{|c|c|c|c|}
 		\hline
 		$\rho$ & 0.05 & $L$ & 1000\\ \hline
 		$\omega_s$ & 60 $\text{s}^{-1}$ & $N_p$ & 13\\ \hline
 		$\omega_a^0$ & 5 $\text{s}^{-1}$ & $\omega_r$ & 0.005 $\text{s}^{-1}$\\ \hline
 		$\omega_d$ & 1 $\text{s}^{-1}$ & $\omega_c$ & 0.1 $\text{s}^{-1}$  \\ 
 		\hline
 	\end{tabular}
 	\label{table_parameter}
 \end{table}

\textit{List of parameters:} In our simulations, we used the reference set of parameters shown in table \ref{table_parameter}. Different choices of parameters are mentioned in the text.

\begin{figure}[t!]
	\centering
	\includegraphics[width=0.48\textwidth]{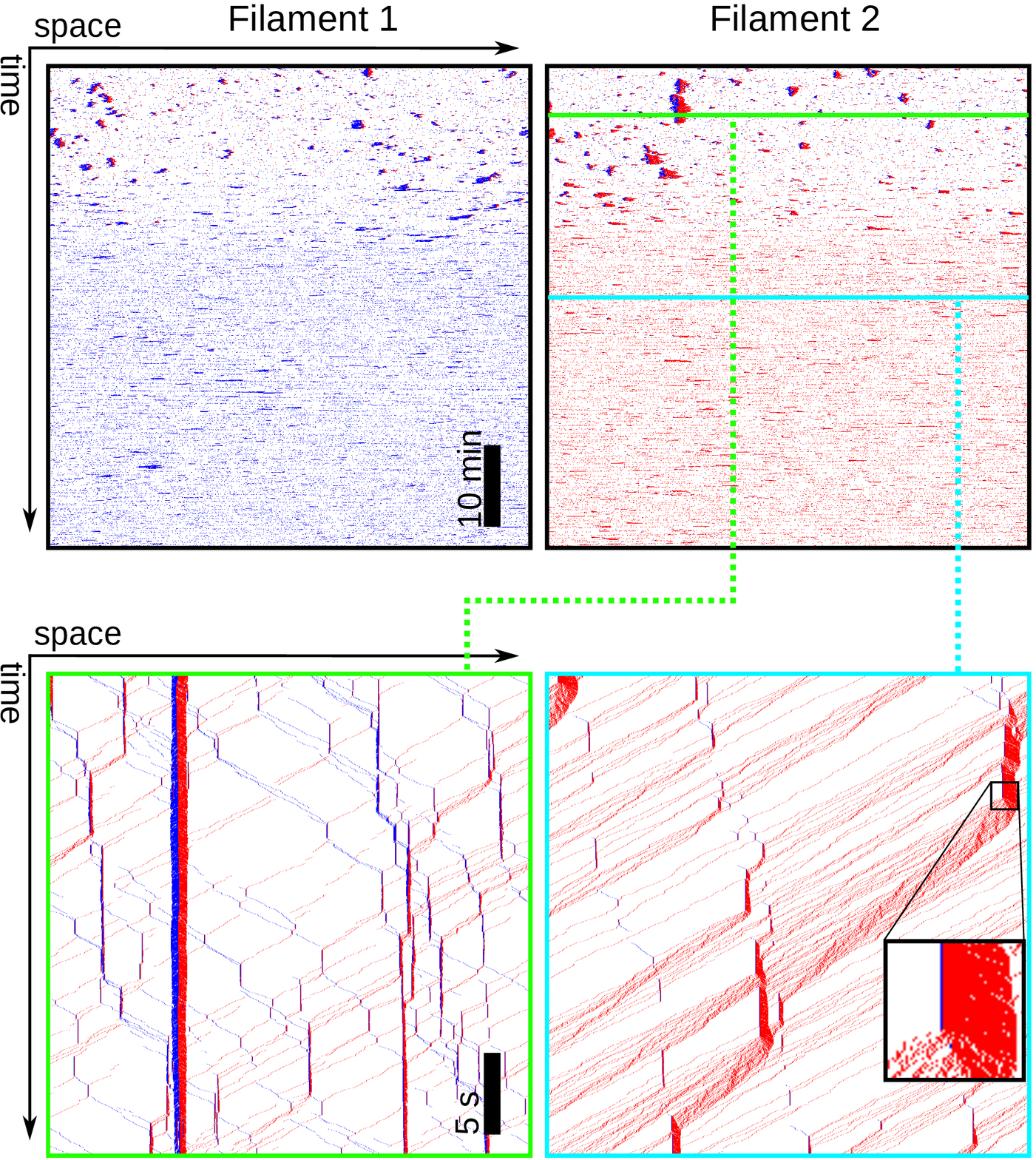}
	\caption{Top: Kymographs in a two-lane system of length $L=1000$, density $\rho=0.08$ and modification $\mu=6$. We studied the time-evolution of the system during the first 
			hour (real time) after initialization in a random configuration and with a neutral floor-field. Bound plus-particles are  blue, minus-particles red and empty space or unbound particles white. The scale car shows a time interval of 10 minutes. The green and cyan line mark 30 s intervals which are shown in the bottom row.
			Bottom: 30 second intervals in a higher resolution show examples of the symmetric state (left) and the asymmetric state (right). The inset shows that typically only a single plus-particle blocks the runs of minus-particles temporally. The scale bar shows an interval of 5 seconds. }
	\label{Sup1}
\end{figure}

\textit{Kymographs:} In order to initialize our simulations, particles are randomly distributed in the unbound state on both filaments. In the top row of Fig.\,\ref{Sup1}, we show a kymograph corresponding to a typical time-evolution of the system over one hour real time. Shown are bound particles of filament 1 ad 2 in a long time interval in order to observe the system transitioning between the two following states. After an initial symmetrical state with clustering, particles distribute asymmetrically between the two filaments. The result is a clear majority of plus- (minus) particles on filament 1 (2). We then call the filament with a plus (minus) majority plus-lane (minus-lane). Also the floor-field clearly breaks symmetry in the same way. This lane-formation was not observed for neutral affinity modification (not shown).

In the bottom row of Fig.\,\ref{Sup1} we show parts of the kymographes in a higher time resolution. Intervals of 30 s are shown from the kymograph of filament 2. The left figure belongs to the early phase in which the system is in a  symmetric state where immobile clusters drastically reduce the flux. The right figure shows the system in the asymmetric state where only a few plus-particles are present on the minus-lane. Minus-particles perform in coordinated runs, only temporary blocked by plus-particles, which results in a more efficient transport state.

\begin{figure}[t!]
 	\centering
 	\includegraphics[width=0.48\textwidth]{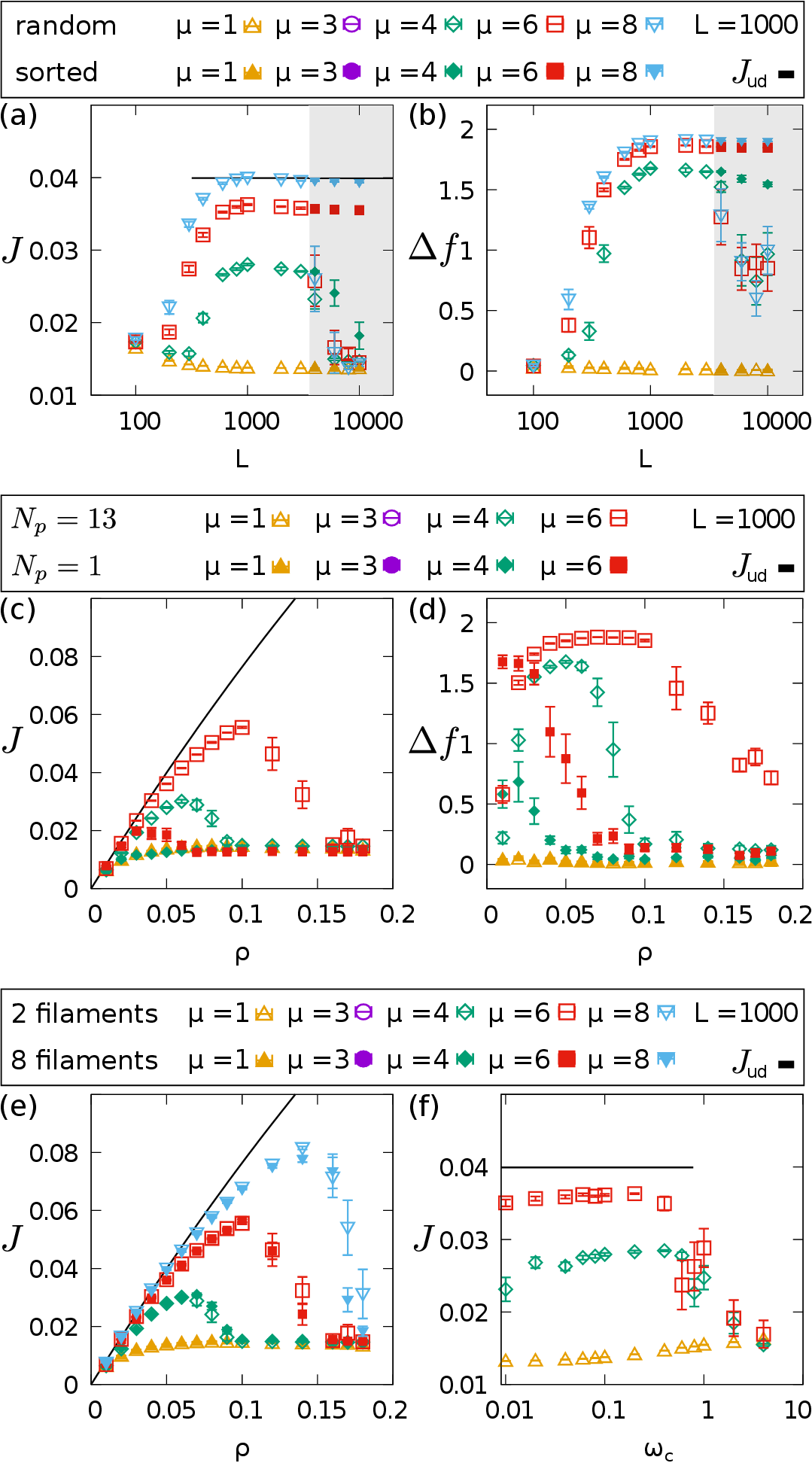}
 	\caption{Flux $\langle J \rangle$ \textbf{(a)} and asymmetry $\Delta f$ \textbf{(b)} under variation of the system length $L$. Simulations did run for 3 hours simulated time, measurements started after 2 hours. Simulations which were initialized  in random configurations are marked as open symbols, simulation data from asymmetric initial conditions are represented by filled symbols. The black line shows $J_{\text{ud}}$ and the shaded area marks the length regime of large error bars and the second set of simulations started in sorted initial conditions.  Model comparison in the fundamental diagram \textbf{(c)} and the floor-field asymmetry \textbf{(d)} between different numbers of sub-states. The reference system with 13 sub-states is given by open symbols, a lattice of only one sub-state by filled symbols. \textbf{(e)} Fundamental diagram for the reference system of 2 filaments (open symbols) compared to a system of 8 filaments (filled). \textbf{(f)} Variation of the filament coupling rate $\omega_c$ for different $\mu$ influences the particle flux. The reference value used in the main text is given by $\omega_c=0.1$.}
 	\label{Sup2}
 \end{figure}

\textit{Length dependency:} Beside the flux and symmetry dependencies on the density $\rho$ and the affinity modification feedback $\mu$, we study the influence of the system length. In Fig.\,\ref{Sup2}(a) and (b) we plot $J $ and $\Delta f$ for a fixed density $\rho=0.05$ under variation of $L$. On the one hand, lane-formation needs a minimum length so that enough particles are involved in the system i.e. the sharper transition in Fig.\,3 in the main text for $L=1000$ than $L=300$ is consistent with the not reached maximum value for $L=300$ in Fig.\,\ref{Sup2} (a) and (b). On the other hand, stable lanes were not able to form within our simulation of 3 hours simulated time for $L\geq 4000$, represented by the large error bars in this regime. For large systems it takes very long  to reach the stationary state from random initial conditions so we check whether asymmetric 
initial conditions are stable and lane-formation persists in the stationary state.  We add filled symbols for simulations started in asymmetric conditions to the open symbols for the reference start in symmetric conditions. Here, the system remains in the asymmetric state when high affinity modification was implied but produced traffic jams and lost asymmetry without the modification. Thus, we expect the system to self organize even for larger system sizes in the stationary state.

\textit{Number of sub-states:} We now check our model calibration in Fig.\,\ref{Sup2} (c) and (d) by comparing two different  numbers of sub-states $N_p=13$ (empty symbols) and $N_p=1$ (filled symbols). The number influences floor-field resistance against single particle induced changes for sub-states and leads to a majority effect. We compare the model to a version with only a single (sub-)state per site. This means, that each particle stepping to a given neutral site sets its affinity, i.e. $f_i=s_i$. This complete modification of the local floor-field by a single particle is not in agreement with experimental results which report a curvature of MTs at low kinesin concentrations corresponding to a partial excitation of the protofilaments [19, 25]. 
In our model the average of $N_p=13$ sub-states takes this collective effect into account 
and determines the local preference from an average of all local-substates $f_i$.
A lattice with only one sub-state shows similar but quantitatively smaller flux enhancement. In case of $\mu=4$, there is no improvement compared to the neutral system. If $\mu=6$, enhancement is visible but only for densities up to $\rho=0.03$ before the flux breaks down. This behavior is also reflected in the floor-field in panel (d). Here, the asymmetry exhibition is shifted towards lower densities and is in case of $\mu=4$ not as strong as for $N_p=13$. For such small densities, the flux is close to $J_{\text{ud}}$ so that the enhancement in the flux is hardly noticeable.

\begin{figure}[t!]
	\centering
	\includegraphics[width=0.48\textwidth]{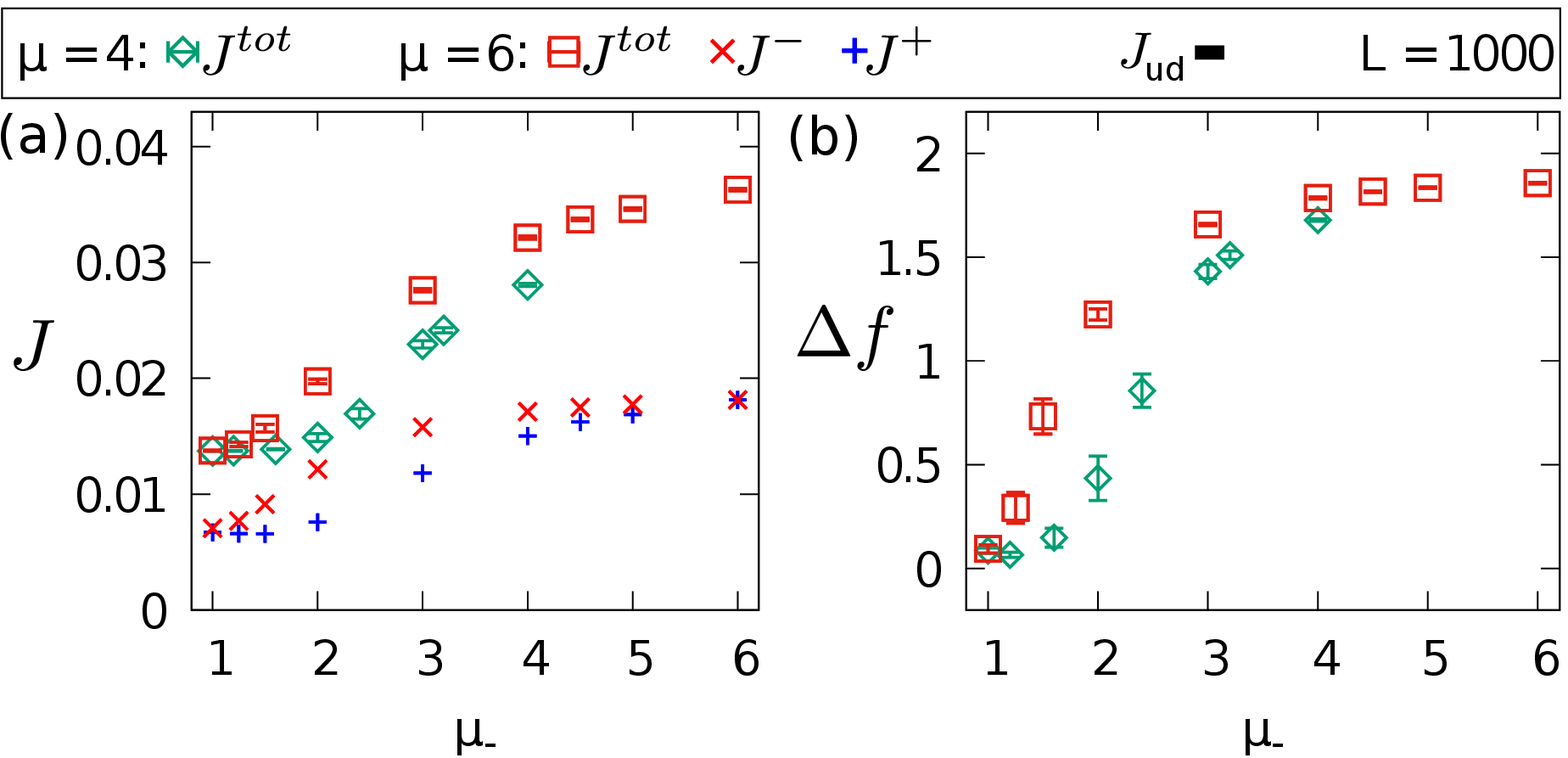}
	\caption{Total particle flux $J^{tot}$ \textbf{(a)} and floor-filed asymmetry \textbf{(b)} in case of asymmetrical motor affinity modification factors $\mu_+$ and $\mu_-$. We fix $\mu_+=\mu$ and vary $\mu_-$ in $[1,\mu]$. For $\mu=6$, fluxes are shown for plus- and minus-particles separately.} 
	\label{Sup3}
\end{figure}

\textit{Number of filaments:} An increasing of the number of lanes does not further improve the flux as it can be seen in panel (e) where filled symbols belong to a system of $N_L=8$ lanes, being in good agreement with the two track version (open symbols).

\textit{Coupling rate $\omega_c$:}
We also investigated the influence of the coupling rate $\omega_c$ which determines the amount of interaction between filaments. The actual value of the coupling 
rate has not been established in experiments.  In all simulations we use the value $\omega_c= 0.1\,\,\omega_d$ as given in table\,\ref{table_parameter}. In Fig.\,\ref{Sup2} (f), the system supports lane-formation and enhanced flux for weakly coupled lanes. The flux breaks down if the coupling rate is of the order of attachment and detachment rate. This breakdown is related to traffic jams on both filaments, located at similar positions. The strong density correlation prevents traffic jams from resolution because motors cannot escape from crowded areas due to exclusion. Strongly coupled filaments cannot organize themselves into lanes anymore. It follows that tracks have to be weakly coupling for the mean-field assumption, which is consistent with studies about unidirectional transport on coupled tracks [42].

The idea of weak coupling is consistent with the biological findings that motor-cargo complexes have a very low diffusivity in the crowded cytosol of the axon \textit{in vivo}. Hence, it makes da difference whether particles attach to a close MT  or to a neighboring MT which is further away.

\textit{Asymmetric motor affinity modification:} 
We investigate if symmetric affinity modification parameters are needed to generate symmetry-breaking in our system. For that, we distinguish $\mu$ for plus-particles $\mu_+$ and minus particles $\mu_-$. For a given $\mu_+=\mu$, we decrease affinity of minus particles down to a neutral setting $\mu_-=1$ and show fluxes and the floor-field asymmetry in Fig.\,\ref{Sup3}.

We can see that no symmetry between plus and minus particle affinity modification is needed for having an enhanced flux and asymmetric floor-fields. However, without a slightly modified affinity of minus- particles, we cannot see any enhancement of the flux no matter if binding of plus-particles is modified or not. This result is consistent with mean-field analysis for $\mu_+=\mu$ and $\mu_-=1$. Eq.\,\ref{Eq:mean-field-simple1} to Eq.\,\ref{Eq:mean-field-simple3} simplify for minus-particles, i.e.
	\begin{align}
	\frac{\partial \rho_t^-}{\partial t}  = \ & \omega_a^0 \rho_u^- - \omega_d \rho_t^- \\	
	\frac{\partial \rho_u^-}{\partial t} = \ & \omega_d\left(\rho_t^- + \rho_b^- \right)  -2\omega_a^0 \rho_u^- \\
	\frac{\partial \rho_b^-}{\partial t} = \ & \omega_a^0  \rho_u^- - \omega_d \rho_b^-.	
	\end{align}
These equation system has only a symmetrical solution. Using this result  in the equations for plus-particles leads exclusively to symmetrical solutions for plus-particles, too. Hence, no lane-formation is found in a mean-field model if the attachment modification applies only to one of the two particle species.

In Fig.\,\ref{Sup3} (a), the plus and minus flux $J^{\pm}$ is not identical for all $\mu_-$. By having only a low affinity modification for minus-particles, it is the plus-particles which cannot produce flux more than in the neutral case. On the filaments, the plus density is concentrated to one filament but minus-particles are located on both. Hence, minus-particles on the minus-lane are rather free to move but most plus-particles are blocked by the minus-particles on the plus-lane. By increasing the minus affinity modification, this gradually changes 
until  the fluxes of the two particle species are balanced.